# A Deep Learning Framework for Short-term Power Load Forecasting

Tinghui Ouyang, Yusen He, Huajin Li, Zhiyu Sun, and Stephen Baek

*Abstract* — The scheduling and operation of power system becomes prominently complex and uncertain, especially with the penetration of distributed power. Load forecasting matters to the effective operation of power system. This paper proposes a novel deep learning framework to forecast the short-term grid load. First, the load data is processed by Box-Cox transformation, and two parameters (electricity price and temperature) are investigated. Then, to quantify the tail-dependence of power load on the two parameters, parametric Copula models are fitted and the threshold of peak load are computed. Next, a deep belief network is built to forecast the hourly load of the power grid. One year grid load data collected from an urbanized area in Texas, United States is utilized in the case studies. Short-term load forecasting are examined in four seasons independently. Day-ahead and week-ahead load forecasting experiments are conducted in each season using the proposed framework. The proposed framework is compared with classical neural networks, support vector regression machine, extreme learning machine, and classical deep belief networks. The load forecasting performances are assessed by mean absolute percentage error, root mean square error, and hit rate. Computational results confirm the effectiveness of the proposed data-driven deep learning framework. The prediction accuracy of both day-ahead forecasting and week-ahead forecasting demonstrate that the proposed framework outperforms the tested algorithms.

*Index Terms*—Load forecasting, tail dependence, deep belief network, Copula model,

## I. INTRODUCTION

RESEARCH for developing a more reliable, efficient, and cost-effective power grid network has attracted significant attentions. For example, smart grid technologies help utilities to speed outage restoration after major storm events and reduce the number of affected customers. Newly developed configuration models are offering more energy supply from renewable energy sources including wind farm and photovoltaic power stations. Power demand (e.g. grid power, load, and etc.) has been widely forecasted to price the electricity to optimize the energy consumption. Among these, short-term power load forecasting is essential to the system's reliability and economic development. Inaccurate load forecasting harms the scheduling and planning of power systems. For example, an overestimation of electricity demand will lead to a conservative operation which may cause units supplying excessive energy.

Underestimation of power load causes unmet demand which forces the system to operate in a vulnerable region to the disturbance [1]. Therefore, the accurate power load forecasting guarantees the safe and stable operation of power system.

In the published literatures, power load forecasting methods have been vigorously studied. According to the prediction models in short-term power load forecasting, they are generally divided into two classes, traditional statistical models and advanced data-driven models. Traditional statistical models are always built by simple regression functions. For example, time series models are typical short-term prediction models in power load forecasting, e.g., autoregressive moving average (ARMA), generalized autoregressive conditional heteroscedasticity (GARCH), and so on. The load demand of a specific power system in Greece was forecasted by traditional ARMA model in [3]. Furthermore, the ARIMA model which is improved by integral functions was applied for prediction of power load in [5]. The GARCH model was introduced for load forecasting in [4]. Quantile regression with kernel-based methods are also reported in [6]-[7]. Apart from these traditional statistical models, data-driven models which are built by artificial intelligence algorithms are more suitable to study the nonlinear relationships. These methods were also widely employed to improve the accuracy of power load forecasting. For example, the support vector machine (SVM) and ant colony optimization algorithms were combined in load forecasting in [8]. The neural networks (NN) were usually applied in short-term power load forecasting [9]. Extreme learning machine (ELM) was an advanced algorithm which is used in prediction areas recent years with decent computational results [10].

Analyzing these methods in previous literatures, there are some problems affecting the accuracy of short-term power load forecasting. For traditional statistical models, which are built by simple regression functions, they are not effective to express the non-linear prediction relationship. The under fitting is also the most common deficiency with comparatively smaller forecasting accuracy. For data-driven models, they have advantages at expressing complex nonlinear relation. The NN related algorithms are mostly applied to forecast short-term power load with high accuracy. However, with a limited number of hidden nodes considered, the capacities of exploring higher nonlinearities are usually constrained. Overfitting is the major disadvantage of a traditional NN with a large number of hidden nodes. On the other hand, peak load is considered as an important factor impacting the reliability of the grid network in both day-ahead load forecasting and week-ahead load forecasting [13]. Even with high forecasting accuracy, any underestimation of the peak load may result power outage. In certain instances, the forecasting of the weekly peak load is the objective of short-term forecasting as it is the most important load during a given time interval [2]. Operation decisions of a

Tinghui Ouyang is with Department of Electrical and Computer Engineering, University of Alberta, Edmonton, Alberta, T6G2R3, Canada (touyang@ualberta.ca)
Yusen He, Zhiyu Sun, and Stephen Baek are with the Department of Mechanical and Industrial Engineering, University of Iowa, Iowa City, IA, 52242, U.S. (yusen-he@uiowa.edu, zhiyu-sun@uiowa.edu, stephen-baek@uiowa.edu)
Huajin Li is with the State Key Laboratory of Geohazard Prevention and Geoenvironment Protection, Chengdu University of Technology, Chengdu, 610059, China (huajinlee@yahoo.com)

power system, e.g. economic scheduling of generations, evaluation of system security, energy transaction planning and so on, are based on the results of short-term power peak load forecasting. Therefore, it is also meaningful to improve the accuracy of the short-term peak load forecasting.

Based on the recent studies, a data-driven deep learning framework is proposed to forecast the short-term power load in this paper. The proposed framework makes three main contributions to improve the accuracy of power load forecasting:

1), A pioneer study of applying deep belief network (DBN) into power load forecasting is presented in this study. Since DBN can overcome the deficiency of traditional NN algorithms and produce more promising results with less hidden neurons for two reasons. First, as the power load forecasting containing inputs with high-dimensionality, a DBN can learn to probabilistically reconstruct the inputs data and then detect the feature patterns. Second, several hidden layers within a DBN only contains a limited number of hidden neurons which can avoid overfitting. Hence, the DBN algorithm is popularly applied in recent related studies to improve the forecasting accuracy [11-12].

2), Two parametric Copula models are applied to construct two indicative binary variables to improve peak load forecasting accuracy. Considering the importance of peak load forecasting, threshold parameter namely Value-at-Risk (VaR) is computed from the fitted Copula model and forms the two new variables to indicate the peak load. Since electricity price and temperatures impacts the power load, the parametric Copula models are fitted with real-time electricity price and average hourly temperature versus power load. Forecasting accuracy of the peak load has been significantly improved by using the Copula models.

According to the above contribution, day-head and week-ahead power load is forecasted using a deep belief network (DBN). Industrial case study is experimented and a comprehensive comparative analysis between the proposed framework and the classical data-driven algorithms is conducted. Three evaluation metrics, namely mean absolute percentage error (MAPE), root mean square error (RMSE), and hit rate (HR) are applied to evaluate the forecasting performance.

The main body of this research is organized as follows. Section 2 describes the methodology of Copula theory and the internal scheme of deep belief network (DBN). Section 3 illustrates the details of the proposed framework, including data transformation, copula modeling, training deep belief network, and evaluation metrics. Section 4 presents forecasting results and comparative analysis of three case studies, day-ahead load forecasting, week-ahead load forecasting, and week-ahead peak load forecasting by applying the proposed framework. Section 5 summarizes the conclusions and findings of this research.

## II. METHODOLOGY

The power load in a grid network system exhibits nonlinear and stochastic behaviors with high variability and volatility. In this research, a data-driven framework is proposed to reduce load forecasting errors. The schematic diagram of the proposed framework is presented in Fig. 1.

According to Fig. 1, this framework incorporates the analysis from Copula models with deep belief network. There are mainly three parts by using the proposed framework in power load forecasting. First, data processing is implemented for any data-driven analysis. Considering the difference of parameters on units and magnitudes, The Box-Cox transformation is used for data normalization. Meanwhile, two influential parameters (i.e., electricity price and temperature) are selected to construct a Gumbel-Houggard Copula model to investigate the upper-tail dependence between the power load and the influential parameters. Then, two peak load indicative binary variables are created as peak load indicators via the Value-at-Risk computation from Copula models and are selected as the inputs to the forecasting model. Second, a forecasting model based on deep belief network is constructed to predict power load including layer-wise pre-training, fine-tuning, and structural optimization. Third, performance evaluation of short-term power load forecasting is examined in day-ahead forecasting, week-ahead forecasting, and week-ahead peak load forecasting cases. To discuss the applicability of the established framework, forecasting are performed in four seasons independently. Consequently, the comprehensive comparative analysis between the support vector regression machine (SVR), neural network (NN), extreme learning machine (ELM), and the classical deep belief network are provided.

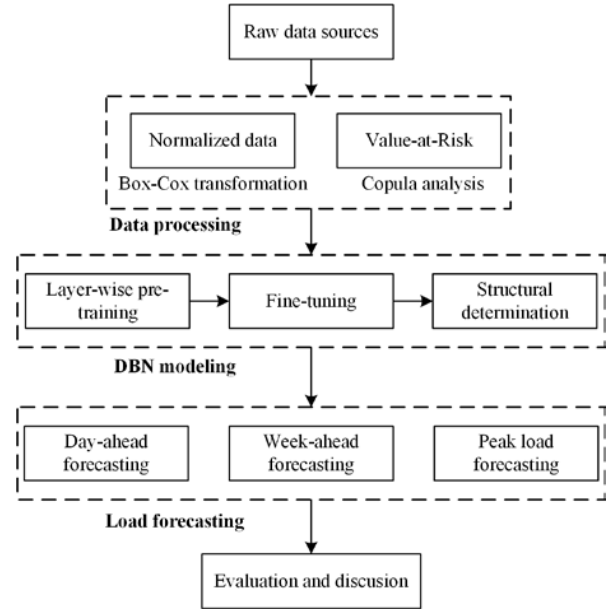

Fig. 1. Schematic diagram of the proposed framework.

### A. Copula Theory

Real-time electricity price and ambient temperature impacts the load of a power system [1]-[2]. For instance, the HAVC systems in buildings would consume more energy under hot weather. Energy consumptions of factories and buildings are influenced by the dynamics of electricity price. To avoid such uncertainties of a grid network, investigation of tail-dependences of power load on temperature or electricity price is essential. In this research, Copula models are introduced to build the relationship between power load verses electricity price and power load verses temperature to escalate modeling accuracy. The Copula model was originally proposed by Sklar [14]-[15] as an *N*-dimensional joint distribution function expressed by an integration of *N* univariate marginal distribution functions and a Copula function. The classical bivariate Copula model can be defined as (1).

$$F(x_1(t), x_2(t)) = C[F_{x_1}(x_1(t)), F_{x_2}(x_2(t))] \quad (1)$$

where: $F_{x1}(x_1(t))$ and $F_{x2}(x_2(t))$ represents the marginal cumulative density functions (CDFs); $x_1$ denotes load; $x_2$ denotes the real-time electricity price or temperature; $F(x_1, x_2)$ is the two-dimensional joint distribution function; and $C(F_{x1}, F_{x2})$ is the Copula function.

*B. Deep Belief Network*

The deep belief network (DBN) has been widely used in the community of deep learning algorithms [16]. It is a fast learning algorithm that locates the optimal solution faster [17]. A classical DBN consists an unsupervised learning subpart using restricted Bolzmann machines (RBMs) as its building blocks and a logistic regression layer for prediction [18].

Restricted Bolzmann machine (RBM) is a stochastic neural network. Layer-wise training is implemented across multiple RBMs in the construction of a deep belief network (DBN). Illustrated in Fig. 2, the RBM contains a layer of binary-valued neurons and a layer of Boolean hidden neurons. There are bidirectional and symmetrical connections between different layers, however, no connection exist between the neurons within the same layer. To learn the probability distribution of the two layers, the learning process of the layer-wise configuration follows its energy function expressed in (2). The probability distribution can be expressed in (3) [19].

$$E(v, h) = -\sum_{i=1}^{n_v} a_i v_i - \sum_{j=1}^{n_h} b_j h_j - \sum_{i=1}^{n_v}\sum_{j=1}^{n_h} h_j W_{j,i} v_i \quad (2)$$

$$P(v, h) = \frac{e^{-E(v,h)}}{\sum_v \sum_h e^{-E(v,h)}} \quad (3)$$

where: $v_i$ is the number of neurons in the visible layer; $h_i$ is the number of Boolean neurons within the hidden layer; $W_{j,i}$ is the weight matrix between the visible layer and the hidden layer; and $a_i$ and $b_i$ are bias vectors for the two layers. Next, the activation functions of each layer are presented in (4)-(5) [19].

$$P(v_i = 1|h) = sig(\alpha_i + \sum_{j=1}^{n_h} W_{j,i} h_j) \quad (4)$$

$$P(h_i = 1|v) = sig(b_j + \sum_{i=1}^{n_v} W_{j,i} v_i) \quad (5)$$

where $sig()$ denotes the sigmoid function.

Single hidden restricted Bolzmann machine (RBM) is incapable of offering sufficient prediction accuracy [20]. A deep belief network (DBN) containing stacked layers of RMBs and a logistic regression layer on the top extracts features from data progressively. Illustrated in Fig .3, the first RBM is pre-trained as an independent RBM using the training data directly. The weight matrix between the visible layer and hidden layer (hidden layer 1) of the first RBM is computed. Next, the output from the first RBM is then treated as the input to the second RBM which also includes a visible layer (hidden layer 1) and a hidden layer (hidden layer 2). Last, a logistic regression layer is stacked on the top and trained using the supervised learning method. After these procedures, the back-propagation (BP) algorithm is used for the fine-tuning process to adjust the parameters of the whole DBN algorithm [21].

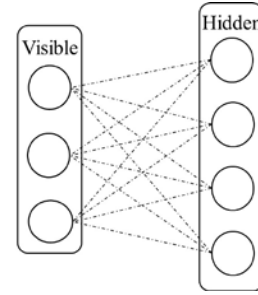

Fig. 2. Restricted Boltzmann machine.

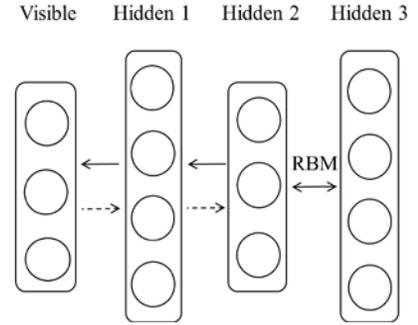

Fig. 3. Deep belief network.

## III. POWER LOAD FORECASTING MODEL

*A. Data Preprocessing*

Based on domain knowledge, the time-series power load data may contain non-normal features (e.g., spikes and fluctuations). These features may deteriorate the forecasting accuracy of the prediction models. One approach to mitigate the negative impact from these features is the normalization of raw data [22]. Box-Cox power transformation is a commonly used method to normalize data [23]. The Box-Cox transformation of time-series power load data is expressed as (6)[24].

$$y_i^{(\lambda)} = \begin{cases} \dfrac{y_i^\lambda - 1}{\lambda} & if \quad \lambda \neq 0 \quad i = 1, \ldots, n \\ \ln y_i & if \quad \lambda = 0 \quad i = 1, \ldots, n \end{cases} \quad (6)$$

where: $\lambda$ is the power parameter; and $n$ is the number of data samples. In this study, the power transformation parameter $\lambda$ is estimated via the Anderson-Darling (A-D) test [25] by using the 10% of the raw data which was sampled randomly.

*B. Gumbel-Hougaard Copula Model*

Preliminary research indicates that power load, temperature, and electricity price exist strong upper-tail dependence [26-27]. In this research, the Gumbel-Hougaard Copula model computes the upper-tail dependence between the power load and the two variables (e.g., temperature and electricity price). The bivariate Copula function of the Gumbel-Hougaard Copula model is expressed in (7). The upper tail dependence parameter of Gumbel-Hougaard Copula model are presented in (8), respectively.

$$C(x_1, x_2) = \exp\{-[(-\ln x_1)^\alpha + (-\ln x_2)^\alpha]^{1/\alpha}\} \tag{7}$$

$$\lambda^{upper} = 2 - 2^{1/\alpha} \tag{8}$$

where: $\alpha$ is the Gumbel-Hougaard Copula model parameter; and $\lambda^{upper}$ denotes the upper tail dependence parameter of a Gumbel-Hougaard Copula model.

*C. Value-at-Risk*

Presented in Section 2.1, the Gumbel-Hougaard Copula model and its properties are introduced in detail. The Copula model parameter $\alpha$ can be estimated by maximum likelihood estimation based on the transformed power load data. Due to the variety of fluctuations and spikes of power load data, an effective statistical quantile estimation of the peak load is crucial. In this research, a tandem indicative variable namely Value-at-Risk (*VaR*), is introduced to increase the prediction accuracy of peak load. The concept of *VaR* was proposed by Jorion [28] as a quantitative measurement metric. It has been widely used as a risk measurement tool for portfolio management of financial assets which contain stochastic movement [29]. In this research, since temperature and electricity price both are stochastic and have impact on the power load [26], two peak load indicative variables (e.g., $I(t)_{temp}$ and $I(t)_{price}$) are constructed based on *VaRs* to improve the peak load forecasting accuracy. With the constructed Copula models, the *VaRs* can be computed by (9)-(10).

$$VaR_p^{temp} = C^{-1}[F_{temp}(x_1(t)), F_{load}(x_2(t))] \tag{9}$$

$$VaR_p^{price} = C^{-1}[F_{price}(x_1(t)), F_{load}(x_2(t))] \tag{10}$$

where $F_{temp}()$, $F_{price}()$, and $F_{load}()$ represent marginal distributions of temperature, electricity price, and power load; $C^{-1}()$ is the inverse function of Gumbel-Hougaard [30] Copula function; $VaR^{temp}$ is the $p^{th}$ upper percentile of bivariate distribution of temperature and power load; and $VaR^{price}$ is the $p^{th}$ upper percentile of the bivariate distribution of electricity price and power load. Hence, the two binary indicative variables of peak load can be expressed in (11)-(12).

$$I(t)_{temp} = \begin{cases} 1 & if \quad temp(t) \geq VaR_p^{temp} \\ 0 & if \quad temp(t) < VaR_p^{temp} \end{cases} \tag{11}$$

$$I(t)_{price} = \begin{cases} 1 & if \quad price(t) \geq VaR_p^{price} \\ 0 & if \quad price(t) < VaR_p^{price} \end{cases} \tag{12}$$

where $I_{temp}()$ and $I_{price}()$ are indicative variables of peak load based on temperature and price. In this research, the value of $p$ is set as 0.95 and the two variables are selected as input in the forecasting model.

*D. Layer-wise Pre-training*

Classical neural networks model adopts back-propagation (BP) as the training principle. The major drawback is that the parameters may easily fall into a local optimum rather than a global one [18]. A promising way to alleviate this drawback is to apply a layer-wise pre-training method for parameter initialization.

In this research, the layer-wise pre-training method requires the deep belief network (DBN) to be pre-trained following a stochastic gradient descent method on the object function of each restricted Boltzmann machine (RBM). As introduced in section 2.2, the probability distribution of an RBM is expressed in (5), the objective function is presented in (13).

$$L(a, b, W) = \sum \log P(v, h) \tag{13}$$

where: $a$ is the bias vector of the visible layer; $b$ is the bias vector of the hidden layer; and $W$ is the weight matrix between layers. The gradient descent method indicates that the parameters (e.g., $a$, $b$, $W$) are updated based on the gradients of the objective function (13). The gradients of the probability distribution function are expressed in (14)-(16)[31].

$$\frac{\partial \log P(v, h)}{\partial W_{j,i}} = \langle v_i h_i \rangle_{P(h|v)} - \langle v_i h_i \rangle_{recon} \tag{14}$$

$$\frac{\partial \log P(v, h)}{\partial a_i} = \langle v_i \rangle_{P(h|v)} - \langle v_i \rangle_{recon} \tag{15}$$

$$\frac{\partial \log P(v, h)}{\partial b_j} = \langle h_i \rangle_{P(h|v)} - \langle h_i \rangle_{recon} \tag{16}$$

where $<>_{P(h/v)}$ is the expectation of the conditional distribution with respect to the input raw data; $<>_{recon}$ is the expectation of the $i^{th}$-step reconstructed distribution. The computation of the expectation regarding the original distribution is efficient and accurate. However, computing the expectation of reconstructed distribution within the deep belief network (DBN) is challenging due to the computational complexity. To solve this problem, contrastive divergence (CD) [32] was proposed to obtain the expectation of the reconstructed distribution through alternating Gibbs sampling [33]. Hence, in this research, the updating rules can be formulated in (17)-(19).

$$W_{i+1} = W_i + \eta(\langle v_i h_i \rangle_{P(h|v)} - \langle v_i h_i \rangle_{recon}) \tag{17}$$

$$b_{i+1} = b_i + \eta(\langle v_i \rangle_{P(h|v)} - \langle v_i \rangle_{recon}) \tag{18}$$

$$a_{i+1} = a_i + \eta(\langle h_i \rangle_{P(h|v)} - \langle h_i \rangle_{recon}) \tag{19}$$

where: $\eta$ is the learning rate; $<>_{P(h/v)}$ is the expectation of the conditional distribution of the input raw data; and $<>_{recon}$ is the expectation of $i^{th}$ reconstructed distribution obtained by alternating Gibbs sampling.

*E. Fine Tuning*

Based on the layer-wise pre-training approach, all parameters of the DBN algorithm are initialized. Adjustment of these parameters in a supervised manner is conducted until the loss function of the DBN reaches its minimum. In this paper, back-propagation (BP) algorithm is applied for the fine-tuning process. All parameters are updated from the top to the bottom resulting reduced forecasting errors. The details of the BP based fine-tuning process is expressed as a flow chart in Fig. 4.

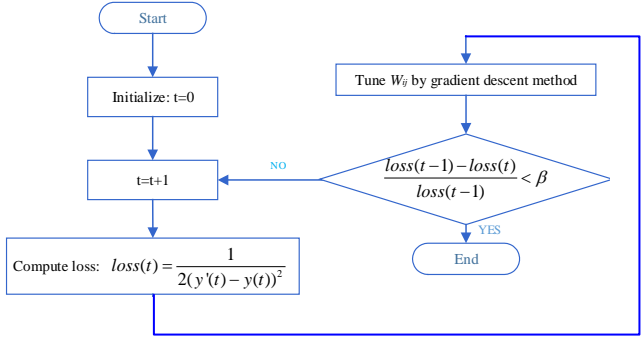

Fig. 4. Back-propagated fine-tuning process.

$$MAPE = \frac{1}{N}\sum_{i=1}^{N}\left|\frac{o_j - t_j}{t_j}\right| \quad (20)$$

$$RMSE = \sqrt{\frac{1}{N}\sum_{j=1}^{N}\left\|o_j - t_j\right\|^2} \quad (21)$$

$$HR = \frac{1}{N}\sum_{i=1}^{N}I(i) \quad (22)$$

where: $o_j$ is the $j^{th}$ predicted power load; $t_j$ is the $j^{th}$ actual power load; $N$ denotes total number of data samples; and $I()$ is the indication function expressed in (23).

$$I(i) = \begin{cases} 1 & \text{if } \left|\frac{o_j - t_j}{o_j}\right| \leq 0.07 \\ 0 & \text{if } \left|\frac{o_j - t_j}{o_j}\right| \geq 0.07 \end{cases} \quad (23)$$

where: $t$ is the number of iteration steps; $y'(t)$ is the predicted value; $y(t)$ is the actual value; $\beta$ is the threshold to ensure that the DBN would not miss any slight update of DBN parameters. When assigned a large value of $\beta$, it is cost infinitive time to obtain an optimal DBN. When assigned an extreme small value, it is likely to result non-convergence. Hence, in this research, $\beta$ is initially set as 0.01. Applying the fine-tuning process iteratively, the optimal values of all parameters are obtained.

*F. Structure Optimization*

With the optimization of model parameters (e.g., $W$, $a$, $b$), the structure of the developed deep belief network (DBN) also needs to be optimized to provide excellent forecasting accuracy. The numbers of hidden neurons hidden layers need to be determined scientifically. In this research, we perform the preliminary structural determination of the DBN based on the randomly selected 10% of raw data. The DBN algorithm is initialized with one hidden layer at first. Hidden-layer neuron numbers ranging from 2 to 40 are examined during the initialization process. Illustrated in Fig. 5(a), the mean absolute percentage error (MAPE) reaches its minimal when there are 30 hidden neurons. Next, the DBN algorithm is initialized with a single hidden layer with 30 hidden neurons contained and the number of hidden layers are increased from 1 to 6 iteratively. Illustrated in Fig. 5(b), the MAPE reaches its minimum when there are three hidden layers. Hence, for the experiments in the case studies, the structure of the deep belief network is 14-30-30-30-1 for the proposed Copula-DBN framework.

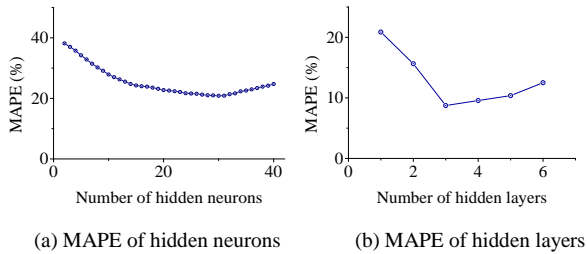

(a) MAPE of hidden neurons   (b) MAPE of hidden layers

Fig. 5. MAPE of neurons and layers.

*G. Performance Evaluation*

To assess prediction accuracy of the proposed framework, three widely used performance evaluation metrics are computed: the mean absolute percentage error (*MAPE* (20)), the root mean square error (*RMSE* (21)), and the hit rate (*HR* (22)).

## IV. CASE STUDIES

One year hourly load data collected from Texas, United States in the year of 2016 is utilized for the case studies. Based on domain knowledge, input variables include: temperature, electricity price, humidity, barometric pressure, wind speed, T-1h power load, T-24h power load, T-48h power load, T-72h power load, T-96h power load, T-120h power load, and T-144h power load respectively. The data pre-processing step include removing invalid values from the dataset is conducted at first.

*A. Tail-dependence Modeling*

With the removal of invalid data, Box-Cox transformation is applied for data normalization. Three normality evaluation metric namely Anderson-Darling test, Jarque-Bera test, and Lilliefors tests are selected to validate the normality of the data after Box-Cox transformation. The computational results are presented in Table I. Since all p-values are larger than 0.05, the normality of the transformed dataset is confirmed. By employing Box-Cox transformation, all values of parameters are normalized and fitted into the Copula models.

In this research, the Gumbel-Hougaard Copula models fit the upper-tail dependence between power load versus real electricity price and power load versus temperature. The default threshold of significance is set as 0.05 and the model parameters are estimated through maximum likelihood estimation. The upper-tail dependence parameter between the power load and the temperature is 3.52. Meanwhile, the upper-tail dependence parameter between the power load and the electricity price is 1.19 respectively. Strong upper-tail dependence of the power load on the temperature and the electricity price have been shown in Fig. 6 significantly. Hence, Value-at-Risk (*VaR*) can be computed as the threshold for peak load indication. The computed *VaRs* at $95^{th}$ percentile are expressed in Table II.

TABLE I. NORMALITY TEST OF PARAMETERS

| Parameters | p-value | | |
| --- | --- | --- | --- |
| | Anderson-Darling | Jarque-Bera | Lilliefors |
| Temperature | 0.84 | 0.57 | 0.82 |
| Electricity price | 0.66 | 0.43 | 0.59 |
| Humidity | 0.99 | 0.85 | 0.99 |
| Barometric pressure | 1.00 | 1.00 | 1.00 |
| Wind speed | 0.24 | 0.08 | 0.21 |
| Power load (T-1h) | 0.78 | 0.69 | 0.73 |
| Power load (T-24h) | 0.56 | 0.46 | 0.51 |
| Power load (T-48h) | 0.94 | 0.80 | 0.91 |
| Power load (T-72h) | 0.95 | 0.76 | 0.86 |
| Power load (T-96h) | 0.65 | 0.46 | 0.61 |
| Power load (T-120h) | 0.94 | 0.80 | 0.89 |
| Power load (T-144h) | 0.82 | 0.63 | 0.74 |

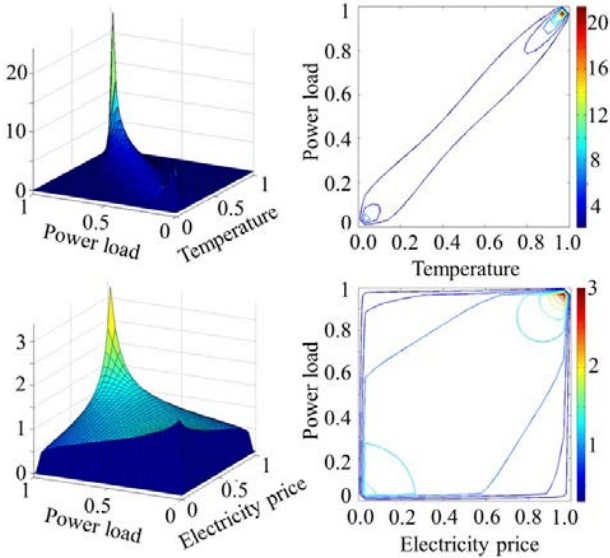

Fig. 6. Fitted Gumbel-Hougaard Copula models.

TABLE II. COMPUTED VALUR-AT-RISKS BASED ON GUMBEL-HOUGAARD COUPLAS

| Power load vs temperature | | Power load vs electricity price | |
| --- | --- | --- | --- |
| Upper-tail parameter | $VaR^{Temp}_{0.95}$ | Upper-tail parameter | $VaR^{Price}_{0.95}$ |
| 3.52 | 92 | 1.19 | 18 |

*B. Day-ahead Power Load Forecasting*

Hourly power load of an urbanized area in Texas, U.S. in the year of 2016 is selected for day-ahead forecasting. The proposed Copula-DBN framework is compared with classical neural networks (NNs), support vector regression machine (SVR), extreme learning machine (ELM), and deep belief network (DBN). The numerical experiments are conducted in four different seasons (e.g., spring, summer, fall, winter). In each season, the one week data is selected as training dataset and the following one day data is selected as validation dataset. For both classical DBN and proposed Copula-DBN framework, the structure is set as 14-30-30-30-1. Layer-wise pre-training and fine-tuning are both implemented to optimize the parameters within the DBN structure. The forecasted day-ahead load is presented in Fig. 7. The experimental results are presented in Table III.

Computational results in Table IV indicates that both DBN and Copula-DBN provides more promising results than other algorithms in day-ahead forecasting. Neural networks and classical DBN provide higher prediction accuracy than SVR and ELM due to their high capacities of capturing non-linearity. However, the prediction errors still exist when there is larger daily variances of power load. The proposed Copula-DBN overcomes this problem attributed to the two indicative peak load indicative variables which reduces the prediction errors at peak load time intervals. For Copula-DBN framework specifically, all performance evaluation metrics (e.g., *MAPE*, *RMSE*, *HR*) are reflecting higher prediction accuracy.

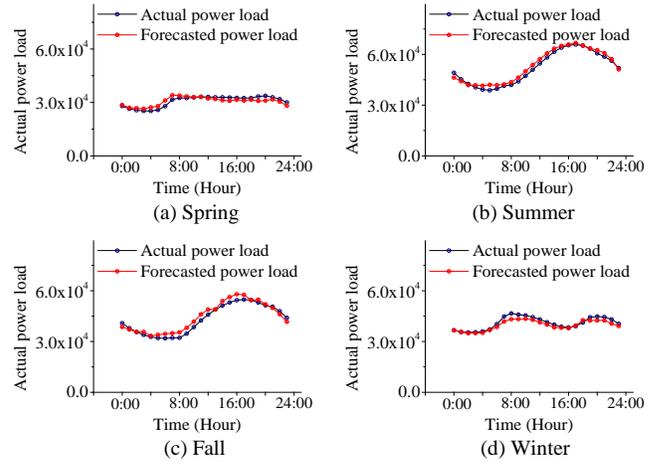

Fig. 7. Day-ahead power load forecasting results.

In Fig. 7, forecasted power load by Copula-DBN framework of four different seasons are illustrated. The daily variances of power load are significantly larger in summer and fall in comparison with daily variances in spring and winter. In spring and winter, the peak load time intervals are between 8:00 a.m and 12:00 p.m. On the other hand, the peak load intervals are between 12:00.p.m and 17:00.p.m in summer and fall. Forecasting the load of next week is discussed in the next section to demonstrate the robustness of the framework.

*C. Week-ahead Power Load Forecasting*

The effectiveness of the proposed Copula-DBN framework in the day-ahead power load forecasting is illustrated. In this section, the week-ahead forecasting as a longer-term forecasting case is selected to examine the framework. In this study, experiments are conducted in four different seasons as well using the past one month data as the training set and the following one week data as the validation dataset. The experimental results have been presented in Table IV and the forecasting results is shown in Fig. 8.

According to Table IV, the values MAPE, RMSE, and HR of Copula-DBN proves its outperformances in comparison with other algorithms. With longer-term prediction horizons, more prediction errors are reduced owing to the modeling capacity of multi-layer deep belief network. In Fig. 8, the majority of the peak loads are captured by the proposed Copula-DBN framework significantly in summer and fall. Computational results confirm the effectiveness of the proposed framework in longer-term forecasting tasks.

TABLE III. PERFORMANCE EVALUATION OF DAY-AHEAD POWER LOAD FORECASTING

| Season | Spring | | | Summer | | | Fall | | | Winter | | |
|---|---|---|---|---|---|---|---|---|---|---|---|---|
| Algorithms | MAPE | RMSE | HR | MAPE | RMSE | HR | MAPE | RMSE | HR | MAPE | RMSE | HR |
| NN | 5.63% | 2568.91 | 87.51 | 6.36% | 1729.13 | 78.14 | 5.53% | 1776.81 | 80.22 | 6.89% | 1921.67 | 71.11 |
| SVR | 5.78% | 2673.34 | 82.34 | 6.52% | 1734.81 | 75.96 | 6.04% | 1893.28 | 76.79 | 7.12% | 2070.74 | 66.39 |
| ELM | 5.93% | 2768.33 | 80.04 | 6.55% | 1735.96 | 75.45 | 5.86% | 1787.45 | 78.27 | 6.96% | 1864.66 | 68.73 |
| DBN | 5.22% | 2288.91 | 88.52 | 5.05% | 1517.62 | 80.52 | 5.34% | 1670.77 | 85.46 | 6.28% | 1750.01 | 76.61 |
| Copula-DBN | **4.95%** | **2031.51** | **89.38** | **4.34%** | **1413.35** | **82.91** | **4.72%** | **1428.24** | **87.33** | **5.34%** | **1646.68** | **78.97** |

TABLE IV. PERFORMANCE EVALUATION OF WEEK-AHEAD POWER LOAD FORECASTING

| Season | Spring | | | Summer | | | Fall | | | Winter | | |
|---|---|---|---|---|---|---|---|---|---|---|---|---|
| Algorithms | MAPE | RMSE | HR | MAPE | RMSE | HR | MAPE | RMSE | HR | MAPE | RMSE | HR |
| NN | 7.37% | 2521.19 | 67.08 | 8.16% | 1593.68 | 59.56 | 7.19% | 2230.02 | 70.02 | 8.25% | 2213.67 | 58.52 |
| SVR | 7.83% | 2734.75 | 64.41 | 8.36% | 1501.12 | 57.16 | 7.65% | 2244.67 | 65.25 | 9.21% | 2350.08 | 55.57 |
| ELM | 7.74% | 2632.31 | 66.63 | 8.49% | 1505.71 | 58.94 | 7.83% | 2150.46 | 63.62 | 9.00% | 2341.87 | 55.89 |
| DBN | 6.99% | 2483.34 | 68.76 | 7.78% | 1479.97 | 60.77 | 6.88% | 2020.01 | 71.73 | 7.99% | 2203.74 | 59.72 |
| Copula-DBN | **6.08%** | **2263.61** | **70.55** | **6.63%** | **1388.84** | **60.82** | **6.21%** | **1917.42** | **72.23** | **7.15%** | **2110.45** | **60.14** |

## C. Week-ahead Peak Load Forecasting

In both day-ahead and week-ahead forecasting cases, the proposed Copula-DBN is proved to be more effective than other algorithms. To further validate the capacity of the proposed Copula-DBN framework, forecasting results in peak load periods are examined specifically. Based on domain knowledge, the majority of the peak load occurs between 8.a.m. and 12.p.m in spring and winter. The peak load in summer and fall mostly occurs between 12.p.m and 5.p.m. Hence, in this section, forecasting performances are evaluated during the peak load period using the computational results from week-ahead forecasting cases. The computational results are presented in Table. V.

With less forecasted data, all values of MAPE, RMSE, and HR in four seasons are smaller than the day-ahead and week-ahead forecasting results. Comparing the other four classical algorithms, the proposed Copula-DBN framework outperforms further contributed by the two peak load indicative variables inputted in the algorithm. Values of MAPE, RMSE, and HR indicate significant difference between Copula-DBN and other algorithms. The comparative analysis confirmed the effectiveness of the proposed Copula-DBN framework in improving peak load forecasting performances.

TABLE V. PERFORMANCE EVALUATION OF WEEK-AHEAD PEAK LOAD FORECASTING

| Season | Spring | | | Summer | | | Fall | | | Winter | | |
|---|---|---|---|---|---|---|---|---|---|---|---|---|
| Algorithms | MAPE | RMSE | HR | MAPE | RMSE | HR | MAPE | RMSE | HR | MAPE | RMSE | HR |
| NN | 2.28% | 1838.47 | 94.44 | 3.77% | 1131.44 | 94.44 | 2.79% | 1728.19 | 94.44 | 3.56% | 1808.83 | 94.44 |
| SVR | 2.59% | 1855.22 | 91.67 | 4.45% | 1335.75 | 91.67 | 3.83% | 1999.63 | 91.67 | 3.71% | 1842.91 | 94.44 |
| ELM | 3.41% | 1954.81 | 88.89 | 4.29% | 1287.66 | 91.67 | 3.66% | 1955.26 | 91.67 | 4.24% | 1963.33 | 91.67 |
| DBN | 2.02% | 1795.63 | 94.44 | 3.82% | 1146.49 | 97.22 | 3.02% | 1788.22 | 94.44 | 3.48% | 1890.66 | 97.22 |
| Copula-DBN | **1.95%** | **1746.02** | **97.22** | **3.74%** | **1122.27** | **97.22** | **2.57%** | **1670.77** | **97.22** | **3.13%** | **1711.14** | **97.22** |

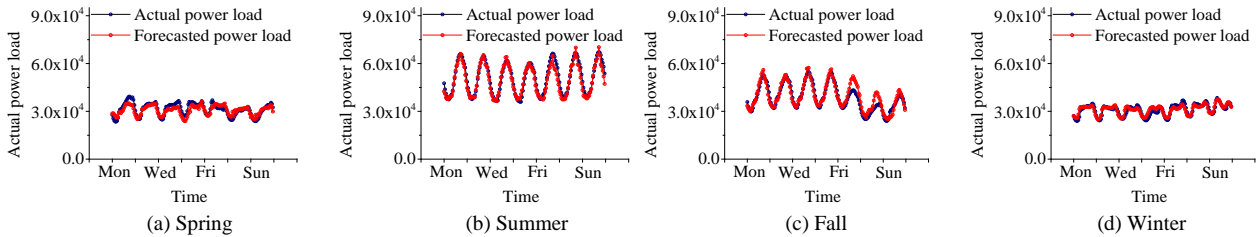

Fig. 8. Week-ahead power load forecasting results.

(a) Spring  (b) Summer  (c) Fall  (d) Winter

## V. CONCLUSION

In this research, a data-driven deep learning framework for power load forecasting has been proposed. Short-term power load forecasting (e.g., day-ahead, week-ahead, and week-ahead peak load) are performed in the case studies. Gumbel-Hougaard Copula model is applied to model the tail-dependence between temperature, electricity price, and power load. Peak load indicative variables are computed from the fitted Copula models. Then, the deep belief network is developed for the day-ahead and week-ahead power load forecasting case studies. A comparative analysis between the support vector regression machine, the classical neural networks, the extreme learning machine, and the classical deep neural networks is conducted using performance assessment metrics.

Computational results supported the effectiveness of the proposed framework for short-term load forecasting. Experimental analysis confirmed the proposed framework outperforms the other algorithms tested to forecast power load. This framework can beneficial for practical short-term scheduling and operations for the grid network.